\documentclass[a4paper]{article}
\usepackage{indentfirst}
\usepackage{hyperref}
\usepackage{multirow}
\usepackage{latexsym,bm,amsmath,amssymb}
\usepackage{graphicx}
\usepackage{epsfig}
\usepackage{subfigure}
\usepackage{cite}
\usepackage{color}
\usepackage[top=1in, bottom=1in, left=1.25in, right=1.25in]{geometry}

\def\no{\nonumber}

\def\pa{\partial}

\def\bea{\arraycolsep .1em \begin{eqnarray}}
\def\eea{\end{eqnarray}}

\def\s0#1#2{\mbox{\small{$ \frac{#1}{#2} $}}}
\def\0#1#2{\frac{#1}{#2}}

\begin{document}

\setcounter{topnumber}{10}
\setcounter{totalnumber}{50}
\title{On the Nature of $X(4260)$}
\maketitle

\begin{center}
{\sc L.~Y.~Dai,$^{\dagger\,,}$\footnote{present address: Center for
Exploration of Energy and Matter, 2401 Milo B. Sampson Lane,
Bloomington, IN 47408,
 USA.
 }\,\,\, Meng Shi,$^\dagger$\,\,\,
Guang-Yi~Tang,$^\dagger$\,\,\,
H.~Q.~Zheng$^{\dagger\,,\star\,,}$\footnote{e-mail address:
zhenghq@pku.edu.cn}}
\\
\vspace{0.5cm}

\noindent{\small{$^\dagger$ \it  Department of Physics and State Key
Laboratory of Nuclear Physics and Technology,
 Peking University, Beijing 100871, P.~R.~China}}\\
\noindent{\small{$^\star$ \it   Collaborative Innovation Center of
Quantum Matter, Beijing, Peoples Republic of China}}
\end{center}
\begin{abstract}
We study the property of $X(4260)$ resonance by
re-analyzing all experimental data available, especially the $e^+e^-
\rightarrow J/\psi\,\pi^+\pi^-,\,\,\,\omega\chi_{c0}$ cross section
data. The final state interactions of the $\pi\pi$, $K\bar K$ couple
channel system are also taken into account.  A sizable
coupling between the $X(4260)$ and $\omega\chi_{c0}$ is found.
The inclusion of the $\omega\chi_{c0}$ data indicates a small value of $\Gamma_{e^+e^-}=23.30\pm 3.55$eV.
\end{abstract}

\section{Introduction}\label{intro}
\begin{figure}
\centering
\includegraphics[width=\linewidth]{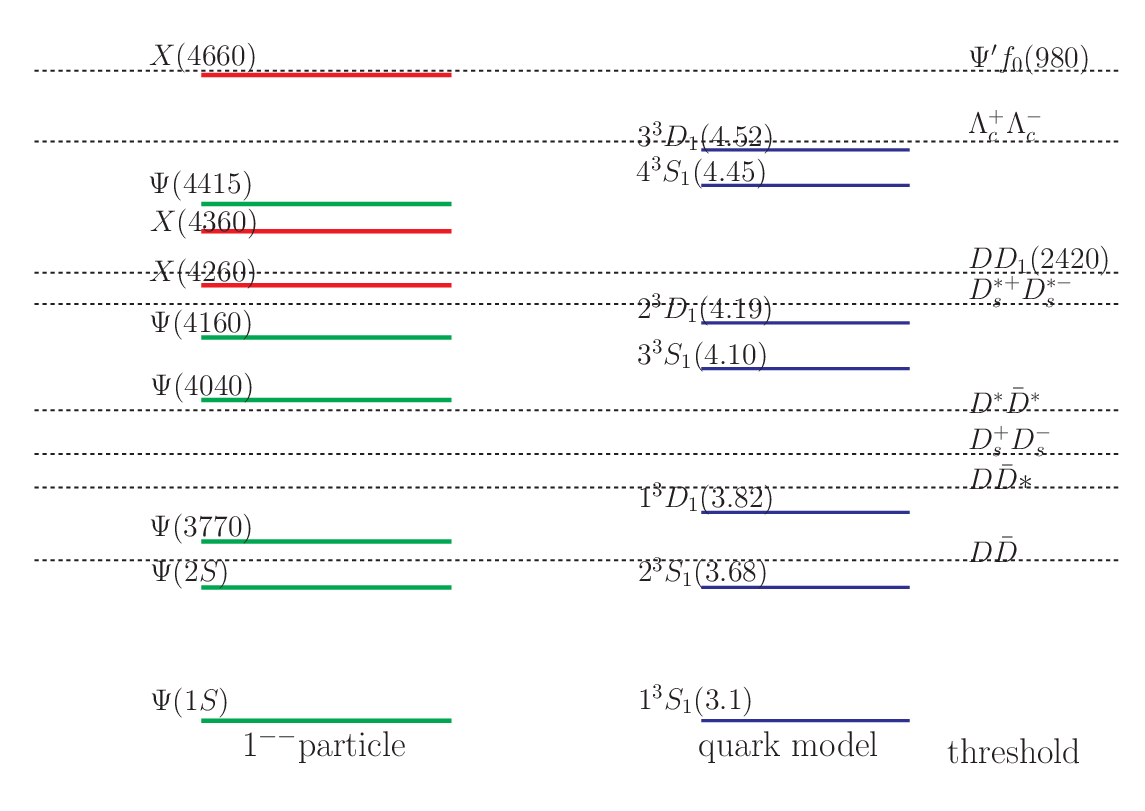}
  \caption{\small X(4260) and nearby resonances from naive quark model calculation\cite{nqm}. }\label{threshold}
 \end{figure}

\begin{figure}
 \centering
 \includegraphics[width=\linewidth]{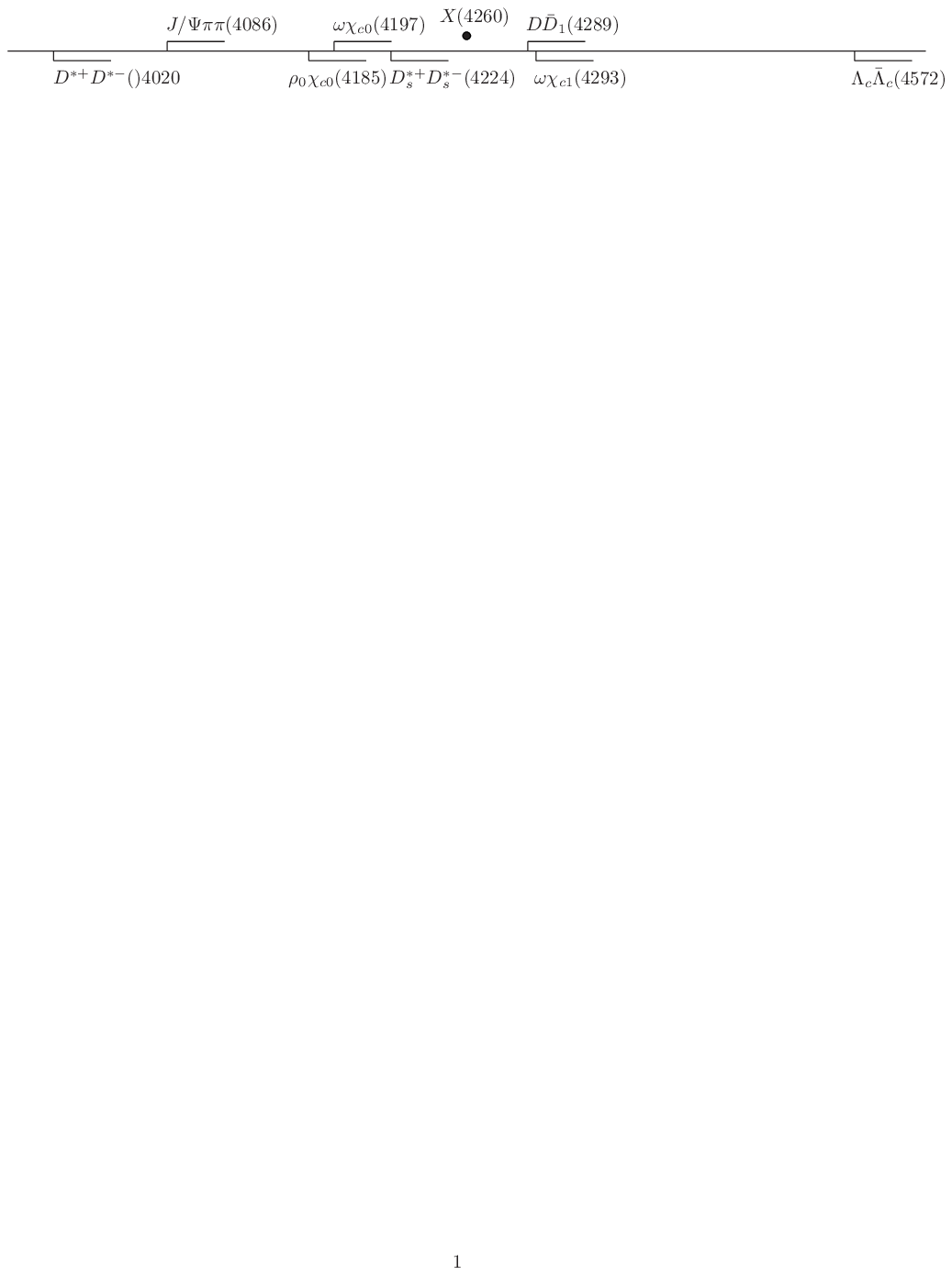}
  \caption{\small Location of X(4260) and nearby thresholds. }\label{X4260threh}
 \end{figure}

 The $X(4260)$ (previously called $Y(4260)$) resonance
 has been found by BaBar Collaboration in initial-state radiation (ISR) process,
 $e^+e^-\rightarrow \gamma_{ISR}J/\psi\pi^+\pi^-$, in
 year 2005\cite{Babary4260}, and  has been confirmed by CLEO\cite{CLEOy4260a} and
 Belle\cite{Belley4260} Collaborations, respectively.
 In  Ref.~\cite{PDG}, the mass and width of this resonance are given with
 $M=4251\pm9$MeV and $\Gamma=120\pm12$MeV, respectively. Furthermore, $\Gamma_{ee}\times\mathrm{Br}(J/\psi\pi\pi)=9.7\pm1.1$eV\cite{Belley4260} or $9.2\pm1.5$eV\cite{BaBar2012}.
In theory aspect, the structure of $X(4260)$ is very interesting,
since it is generally thought that there are not enough unassigned
vector states in charmonium spectrum {(including
the recently reported $Y(4360)$, $X(4630)$/$Y(4660)$ states)}, also the masses  are
inconsistent with naive quark model predictions\cite{nqm} -- the
only such $1^{--}$ states expected up to $4.4$GeV are 1S, 2S, 1D,
3S, 2D and 4S, and they seem to be well established~\cite{review}.
The situation is depicted in Figure~\ref{threshold}.
It is  noticed that above $D\bar{D}$ threshold the number of
$1^{--}$ states given by quark model prediction is inconsistent
with that given by experiments. One tends  to believe that the
discrepancy between the naive quark model prediction and the
observed spectrum is ascribed, at least partially, to the existence
of many open charm thresholds, since the latter will distort the
spectrum. The situation is depicted in Figure~\ref{X4260threh}.
Because of the situation as described above many theoretical papers
have been devoted to the investigation on  $X(4260)$. In
the literature, many models have been made, e.g.,
 $\chi_{c0}\rho^0$ molecule\cite{Y4260molecule1}, $\omega\chi_{c1}$
  molecule\cite{Y4260molecule2},    $c\bar c g$ hybrid
state~\cite{close05,zhusl,0607083,0507119,chengyin2014},
$\Lambda_c\bar\Lambda_c$ bayronium\cite{Y4260molecule3}, $D_1\bar D$
  or $D_0\bar D^*_0$ molecule\cite{Y4260molecule4,Y4260molecule5,zhaoqiang2013},
  non-resonant explanation~\cite{vanBeveren:2010mg,nonresonant}, etc.. Besides, the
  tetraquark state explanation is also very intersting\cite{12102990,13036857,10112818,08083912,08021806,0507062},
  especially when
two resonances, $Z_c(3900)$ and $Z_c(4025)$, are recently found in
$J/\psi\pi$ and $D^{(*)}\bar D^{(*)}$ channels in $e^+e^-$
annihilation near $4.26$GeV by BESIII Collaboration\cite{13035949},
and confirmed by Belle \cite{13040121} and CLEO \cite{13043036}
Collaborations.
However the open charm channels such as $\bar D D^*,D\bar
D^*,D^*\bar D^*$ are not found in the final states of X(4260)
decays\cite{Babar2009,PhysRevD.77.011103,PhysRevLett.98.092001},
making the property of $X(4260)$ to be more mysterious.

 The present authors have also studied the $X(4260)$ issue in the previous edition of the
 present paper (Preprint arXiv:1206.6911v2, herewith denoted as V1).
 Through a careful analysis to experimental data available, it is found that the $X(4260)$ has a sizable
 coupling to $\omega\chi_{c0}$ channel,  but not to other (nearby)
 channels. Inspired by our result, a recent experimental analysis~\cite{omegachic0scp,Ablikim:2014qwy}
 shows that there is indeed a sizable $\omega\chi_{c0}$ final state signal in $e^+e^-$ collision at around $4.26$GeV,
 which hints that the $X(4260)$ may have a large coupling to
 $\omega\chi_{c0}$ --
 though it is not totally clear whether the $\omega\chi_{c0}$ is from the continuum spectrum or the X(4260) resonance, or from both.
 Furthermore, the cross section of $h_c\pi\pi$ channel\cite{Ablikim:2013wzq} is also measured at this energy
 implying that the X(4260) may also couple to it. Because of all these new observations, we have an urge to
 upgrade the work of V1. In the present paper, we continue the
 preceding analysis by including the $\omega\chi_{c0}$ (and also $h_c\pi\pi$) cross section data,
 and we find that the major qualitative conclusion of V1
 still holds, that is $X(4260)$ couples significantly to $\omega\chi_{c0}$  but not to other nearby thresholds.
 Furthermore, we find that the $X(4260)$ resonance is likely to
 maintain a small $e^+e^-$ width, thanks to the new $\omega\chi_{c0}$ data.

 The paper is organized as the following:
 In section~\ref{Jpipi-amplitude} we review on theoretical tools we use in this paper,
  where special emphasis is made on the final state interactions between pions and kaons.
  In section~\ref{Numerical} we
 give a detailed description to our numerical fit program with two scenarios, one does not
 include the $\omega\chi_{c0}$ cross section data \cite{omegachic0scp,Ablikim:2014qwy}, and the other one includes.
  Both of them take into account the effect of the possible $h_c\pi\pi$ decay channel.
   The pole locations of the $X(4260)$ propagator are also searched for.
    Finally  conclusions and physical discussions on the present analysis  is given in section \ref{conclusion}.

\section{Theoretical Discussions on  $e^+e^-\rightarrow J/\psi\pi\pi$}
\label{Jpipi-amplitude}
\subsection{Effective Lagrangian Describing $e^+e^-\rightarrow J/\psi\pi\pi$ Interactions}
Assuming that the $X(4260)$ is a $J^{PC}=1^{--}$ chiral
singlet particle. The transition operator between
photon and $X(4260)$ is as the following:
 \bea\label{eq1}
\mathcal{L}_{\gamma X}&=&g_0 X_{\mu\nu}F^{\mu\nu},
\eea
where we use the anti-symmetric representation $X_{\mu\nu}$ to describe the $1^{--}$ state $X(4260)$, and $F_{\mu\nu}$ denotes the photon field strength.
Notice that in the present notation, one has
 \bea\label{Gammaee}
 \Gamma_{e^+e^-}=\frac{4\alpha}{3}\,\frac{g_0^2}{M_X}\ ,
  \eea
where we have neglected the electron-positron masses.
 For $X(4260)$ decay, the following effective lagrangian is used, which is accurate in the leading order in the expansion in terms of $\pi$ momentum in the center of mass
frame of $\pi\pi$ system:
\bea\label{XJPsipipi}
\mathcal{L}_{X\psi PP}&=&h_1 X_{\mu\nu}\psi^{\mu\nu} <
u_{\alpha}u^{\alpha} > +h_2 X_{\mu\nu}\psi^{\mu\nu}<\chi_+> + h_3
X_{\mu\alpha}\psi^{\mu\beta}<u_{\beta}u^{\alpha}> \ ,
 \eea
where anti-symmetric representation $\psi^{\mu\nu}$ describes $J/\psi$. Up to $O(p^2_\pi)$ level, in Eq.~(\ref{XJPsipipi}) there exist only three
independent interaction terms with coefficients $h_1$, $h_2$ and
$h_3$. Further, $u_\mu =i(u^{+}\partial _\mu u-u\partial _\mu
u^{+})$ and
\begin{eqnarray}
u=\exp\{i\frac{\Phi}{\sqrt{2}F_\pi}\}
\end{eqnarray}
is the parametrization of the pseudo-goldstone octet:
\begin{equation}
\Phi=
 \left( {\begin{array}{*{3}c}
   {\frac{1}{\sqrt{2}}\pi ^0 +\frac{1}{\sqrt{6}}\eta _8 } & {\pi^+ } & {K^+ }  \\
   {\pi^- } & {-\frac{1}{\sqrt{2}}\pi ^0 +\frac{1}{\sqrt{6}}\eta _8} & {K^0 }  \\
   { K^-} & {\overline{K}^0 } & {-\frac{2}{\sqrt{6}}\eta_8 }  \\
\end{array}} \right).
\end{equation}
The chiral symmetry breaking term with coefficient $h_2$ in
Eq. (\ref{XJPsipipi}) reads,
\begin{eqnarray} \chi_+
=u^{+}\chi u^{+}+ u\chi^{+}u,
  \,\,\,\,\,\, \chi=2B_0\,\mathrm{diag}(m_u, m_d, m_s)\ .
 \end{eqnarray}
Parameters $F_\pi$ and $B_0$ can be fixed phenomenologically: $F_\pi
\approx 92.4$MeV and $<0\mid\psi
\overline{\psi}\mid0>=-F^{2}B_{0}[1+O(m_{q})]$. The Eq. (\ref{XJPsipipi})
can also be rewritten in an explicit form,
 \bea
 \mathcal{L}_1&=&\frac{4h_1}{F_{\pi}^2}X_{\mu\nu}F_{}^{\mu\nu}(\pa_{\rho}\pi^+\pa^{\rho}\pi^-+
 \frac{1}{2}\pa_{\rho}\pi^0\pa^{\rho}\pi^0+\pa_{\rho}K^+\pa^{\rho}K^-+
 \pa_{\rho}K^0\pa^{\rho}\bar
 K^0+\frac{1}{2}\pa_{\rho}\eta\pa^{\rho}\eta),\no\\
 \mathcal{L}_2&=&-\frac{4h_2}{F_{\pi}^2}X_{\mu\nu}F_{}^{\mu\nu}(m_{\pi}^2\pi^+\pi^-+
 \frac{1}{2}m_{\pi}^2\pi^0\pi^0+m_K^2K^+K^-+m_K^2K^0\bar K^0+
 (\frac{2}{3}m_K^2-\frac{1}{6}m_{\eta}^2)\eta\eta),\no\\
  \mathcal{L}_3&=&\frac{4h_3}{F_{\pi}^2}X_{\mu\alpha}F_{}^{\mu\beta}
 (\frac{1}{2}\pa_{\beta}\pi^+\pa^{\alpha}\pi^-+\frac{1}{2}\pa_{\beta}\pi^-\pa^{\alpha}\pi^+
 +\frac{1}{2}\pa_{\beta}\pi^0\pa^{\alpha}\pi^0
+\frac{1}{2}\pa_{\beta}K^+\pa^{\alpha}K^-+\frac{1}{2}\pa_{\beta}K^-\pa^{\alpha}K^+\nonumber\\
 &&+\frac{1}{2}\pa_{\beta}K^0\pa^{\alpha}\bar K^0+ \frac{1}{2}\pa_{\beta}\bar K^0\pa^{\alpha}K^0
 +\frac{1}{2}\pa_{\beta}\eta^0\pa^{\alpha}\eta^0)\ .
\eea

\subsection{Kinematics and Tree Level Amplitudes}\label{tree}
\begin{figure}
\vspace{1cm}
 \centering
  \psfig{file=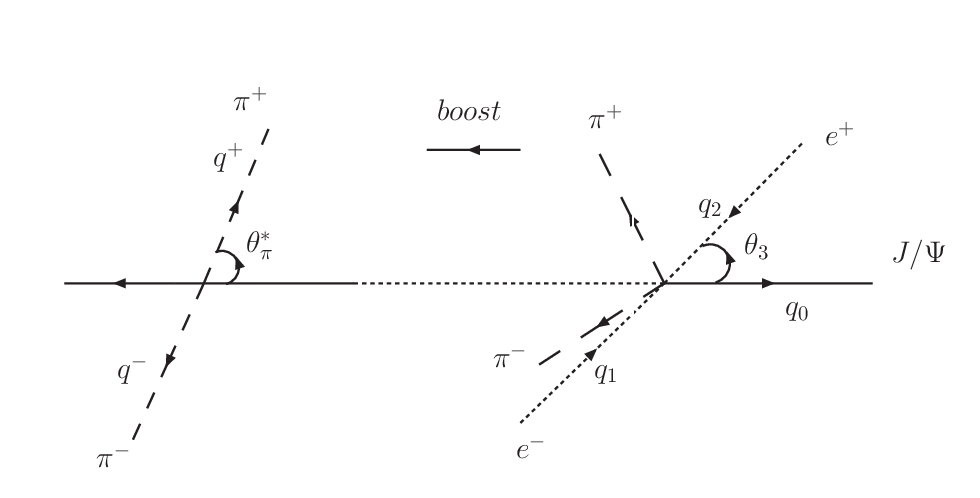,width=12.5cm,angle=-0}
  \caption{A depiction of kinematics.}\label{figg11}
 \end{figure}

We denote the momenta of $e^-$, $e^+$, $X(4260)$, $J/\psi$, $\pi^+$
and $\pi^-$ as $q_1$, $q_2$, $q$, $q_0$, $q^{+}$ and $q^{-}$ respectively,
see in Fig. \ref{figg11}. The polarization of $J/\psi$ is
represented as $\epsilon_{\psi}$, and $k_\pm=q^{+}\pm q^{-}$.
Then one has the following relations:
 \bea &&s\equiv k_+^2,\nonumber\\
      &&k_-^2=-s\rho(s)^2=4m_{\pi}^2-s,\nonumber\\
      &&q_0^2=M_{J/\psi}^2,
   \,\,\, k_+\cdot k_-=0,\nonumber\\
      &&k_+\cdot q_0=\frac{1}{2}(q^2-M_{J/\psi}^2-s),
      \eea
  where $\rho(s)=\sqrt{1-\frac{4m_{\pi}^2}{s}}$.
 The amplitude of $X\to J/\psi\pi^+\pi^-$ process at tree level is
  \bea\label{treeamp}
  i\mathcal{A}^{tree}&=&\frac{i 4 e g_0 }{ M_{J/\psi } F_\pi^2 q^2 D_X(q^2)}
\bar{v}(q_1,s)\gamma_{\lambda} u(q_2,s')\{
 [ 4 h_1\frac{1}{2}(s-2 m_\pi^2) +4 h_2 m_\pi^2 ] (q^{\it{0}}\cdot q \epsilon_\psi^{\lambda}-q\cdot\epsilon_\psi q^{\it{0}\lambda}) \no\\
 &&+ \frac{1}{2}h_3[~-q^{\it{0}}_\alpha q\cdot \epsilon_\psi (k_+^{\lambda}k_+^{\alpha}-k_-^{\lambda}k_-^{\alpha})
   +\epsilon_\psi^{\lambda}q^{\it{0}}_\alpha q_{\beta}(k_+^{\alpha}k_+^{\beta}-k_-^{\alpha} k_-^{\beta})\no\\
    &&  -q^{\it{0}\lambda}q_{\alpha}\epsilon_{\psi \beta}(k_+^{\alpha}k_+^{\beta}-k_-^{\alpha}k_-^{\beta})
   +q^{\it{0}}\cdot q   \epsilon_{\psi\alpha}(k_+^{\lambda}k_+^{\alpha}-k_-^{\lambda}k_-^{\alpha})]
   \},
 \eea
 where $\alpha, \beta, \lambda$ are Lorentz indices and $D_X(q^2)$ is the denominator of the $X(4260)$ propagator which will be discussed latter.

 Following the helicity amplitude decomposition method~\cite{tensor} and choosing
the basis of tensors
\begin{eqnarray}
 \tilde{t}^{(0)}&=&1,\no\\
 \tilde{t}^{(1)}&=&k_-^{\mu},\no\\
  \tilde{t}^{(2)}=k_-^{\mu}k_-^{\nu}-\frac{1}{3}k_-^2\tilde{g}^{\mu\nu}&,&
   (\tilde{g}^{\mu\nu}=g^{\mu\nu}-\frac{k_+^{\mu}k_+^{\nu}}{k_+^2}),
\end{eqnarray}
 it is easy to separate the $S$-wave and $D$-wave components of $\pi\pi$ system. The overall $S$-wave amplitude reads,
 \bea\label{A}
i \mathcal{A}_s^{tree}=&&\frac{i 4 e g_0 }{ M_{J/\psi} F_\pi^2 q^2
D_X(q^2)}\bar{v}(q_1,s)\gamma_{\lambda}
u(q_2,s')\epsilon_{\psi\omega}\{
 [ 4 h_1 \frac{1}{2}(s-2 m_\pi^2) +4 h_2 m_\pi^2 ]
 (q^{\it{0}}\cdot q~ g^{\lambda\omega}-q^\omega q^{\it{0}\lambda}) \no\\
&&+ \frac{1}{2}h_3[-\frac{1}{3}\rho^2(s)q^0_{\lambda}q^{\omega}s +\left(1-\frac{1}{3}\rho^2(s)\right)
 k_+^{\lambda}q^\omega q^{\it{0}}\cdot k_+
 +\frac{1}{3}\rho^2(s)q^0\cdot q~ g^{\lambda \omega}s\no\\
&&-\left(1-\frac{1}{3}\rho^2(s)\right)g^{\lambda\omega}
  k_+\cdot q~ k_+\cdot q^{\it{0}}
  -\frac{1}{3}\rho^2(s)q^0_{\lambda}q^{\omega}s +\left(1-\frac{1}{3}\rho^2(s)\right)
 k_+^{\lambda}q^{{\it{0}}\omega} q\cdot k_+
  \no\\ &&
 +\frac{1}{3}\rho^2(s)q^0\cdot q~ g^{\lambda \omega}s -\left(1-\frac{1}{3}\rho^2(s)\right)k_+^\lambda k_+^\omega q\cdot q^{\it{0}} ] \},
 \eea
whereas the $D$-wave part is then
 \bea
i \mathcal{A}_d^{tree}&=&\frac{i 4 e g_0 }{ M_{J/\psi} F_\pi^2 q^2
D_X(q^2)}\bar{v}(q_1,s)
\gamma_{\lambda} u(q_2,s')\epsilon_{\psi}^{\omega} \no\\
 && \frac{1}{2}h_3\left(t_2^{\lambda\alpha}q^0_{\alpha}q_{\omega}-t_2^{\alpha\beta}g_{\lambda\omega}q^0_{\alpha}q_{\beta}
 +t_2^{\alpha\beta}g_{\beta\omega}q^0_{\lambda}q_{\alpha}
 -t_2^{\lambda\alpha}g_{\alpha\omega}q^{\it{0}}\cdot q \right)\ \no\\
&\equiv& \frac{2eg_0h_3}{F^2_\pi M_{J/\psi}q^2D_X(q^2)}B_d\ .
 \eea
 The standard Breit-Wigner type of X(4260) propagator is parametrized as $D_X(q^2)=M^2_X-q^2-iM_X\Gamma_X(q^2)$, where $\Gamma_X(q^2)$ is the total decay width including partial width of all possible channels that will be discussed latter.

 Since the $D$-wave contribution is proportional to the 4th power of the kinematic factor $\rho(s)$,
 it is highly suppressed comparing with the $S$-wave contribution.
Through numerical studies it is shown that the $D$-wave  contribution is roughly less than $1\%$ of the total decay rate, therefore we will not include it in the fitting process in this work.

\subsection{Final State Interactions(FSI)}\label{FSI}
The tree level amplitude as described in section~\ref{tree} is not
sufficient to describe the $X\to J/\psi\pi\pi$ decay process, since
the $\pi\pi$ system undergoes strong final state interactions (FSI),
especially in $IJ=00$ channel. To include FSI, the following decay
amplitude is proposed \cite{pennisr}:
 \bea\label{MP}
\mathcal { A}_{1}&=&\mathcal {A}^{tree}_{1}\alpha_1(s)
T_{11}(s)+\mathcal {A}_{2}^{tree}\alpha_2(s) T_{21}(s)\
 ,\no\\
 \mathcal {A}_{2}&=&\mathcal {A}_1^{tree}\alpha_1(s) T_{12}(s)+\mathcal {A}_{2}^{tree}\alpha_2(s) T_{22}(s)\
 ,
 \eea
where subscripts 1,2 denote $\pi\pi$ and $K\bar K$ final states,
respectively. For $A_2^{tree}$(the $K\bar K$ amplitude), one only needs to change the $m_\pi$ into $m_K$ in $A_1^{tree}$(the $\pi\pi$ amplitude). Especially  $T_{11}$,  $T_{12}$ and $T_{22}$ represent
$\pi\pi\rightarrow \pi\pi$,  $\pi\pi\rightarrow
 K\bar K$,   $K\bar K\rightarrow K\bar K$ scattering amplitudes,
 respectively. Functions $\alpha_i$ are mild polynomials which place the role to offset
 the `left hand' cuts on the complex $s$ plane
 in amplitude $T$ that would not appear in function $\mathcal{A}$.
Expressions in Eq.~(\ref{MP}) are remarkable in the sense that the
unitarity relations are automatically satisfied. The decay amplitudes $\mathcal{A}_i$, as
an analytic function of $s$, obey:
 \bea
  \mathrm{Im}\mathcal{A}_{1}&=&\mathcal{A}_{1}^*\rho_1T_{11}+\mathcal{A}_{2}^*\rho_2T_{21},\no\\
  \mathrm{Im}\mathcal{A}_{2}&=&\mathcal{A}_{1}^*\rho_1T_{12}+\mathcal{A}_{2}^*\rho_2T_{22}.\label{unitary}
  \eea
 In $\alpha_1(s)$ an additional pole term is added:
  \bea\label{eq2}
  \alpha_1(s) &=&\frac{c^{(1)}_0}{s-s_A}+c^{(1)}_1+c^{(1)}_2\, s +\cdots\ ,
  \eea
  where $s_A$ represents the Adler zero of $T_{11}$. The role of the  pole term
  is to cancel the Adler zero hidden in $T_{11}$ but  not welcome in $A$~\cite{pennisr}.
  An advantage of the pole term in Eq.~(\ref{eq2}) is that, by appropriately choosing coefficient $c^{(1)}_0$
  as $\lim_{s\to s_A}\frac{c^{(1)}_0}{s-s_A}\,T_{11}(s)=1$, it guarantees
 \bea
\mathcal{A}_1= \mathcal{A}_1^{tree} + O(s^2). \eea Up to the leading order
 in $\chi$PT, one finds $c^{(1)}_0=16\pi F_\pi^2$ and
$s_A = m_\pi^2/2$ in Isospin 0 S-wave. However in [1,1] Matrix
Pad\'e amplitude,  $s_A=(0.490-0.008i)m_\pi^2$ and we fix
$c_0^{(1)}=(0.330-0.001i)\text{GeV}^2=(0.779-0.002i)16\pi F_\pi^2$.
\footnote{The Adler zero  moves to the complex plane because the
existence of the left hand cut $(-\infty, 4m_K^2-4m_{\pi}^2]$ of
$T_{22}$, which has been taken into $T_{11}$ due to Matrix Pad\'e
approximation. And so on for that of $c_0^1$.}


\begin{table}[h]
\begin{center}
 \begin{tabular}  {|c|c|c|c|c|c|c|}
 \hline
$L_1$         &      $L_2$    &      $L_3$       &       $L_4$    &
$L_5$        &  $2L_6$+$L_8$  &  $2L_7$+$L_8$  \\ \hline
0.881       &  1.029      &  -3.803         &  0.176     & 1.111     &  1.123 & 0.392  \\
 \hline
 \end{tabular}
 \caption{\label{LECs}
 Low energy constants from couple channel Pad\'e amplitudes.
 Here these parameters are
 refitted and are slightly different from Ref.~\cite{pade}. The unit is $10^{-3}$.}
 \end{center}
\end{table}

In V1, three different representations of $T$ matrices were used for
the fit (coupled channel Pad\'e approximation~\cite{pade}, K-matrix
unitarization~\cite{yumao} and the PKU representation~\cite{pku2} ),
however not much difference was obtained. Therefore we only keep the
 Pad\'e amplitude in this work to perform the
fit to the $\pi\pi$ invariant mass spectrum. In Table~\ref{LECs} we
list the low energy constants\footnote{We noticed that for these
LECs there is a  difference between ours~\cite{pade} ($L_i^{O}$) and
that in the earlier work~\cite{Pelaez02}($L_i^{P}$):
$L_i^{P}=L_i^{O}+\frac{\Gamma_i}{32\pi^2}$. It is because we
calculated in $\overline{MS}$ and theirs in $\overline{MS}-1$.  To
compare with their work we need to transform our LECs into
$L_i^{P}$, which has been incorrectly transferred in~\cite{pade} and we
correct it here.}). For more details about these T matrices, we
refer to the original literature.

\section{Numerical Analysis}\label{Numerical}
\subsection{The Experimental Data and Fit Process}
Once the $X(4260)\rightarrow J/\psi\pi\pi$ amplitude is calculated from Eq.~(\ref{MP}) in above section, one obtains the decay width $\Gamma_{J/\psi\pi\pi}$, the cross section $e^+e^-\rightarrow X(4260)\rightarrow J/\psi\pi\pi$ and the $\pi\pi$ invariant mass spectrum.
 In this subsection we focus on how to write down the correct form of  denominator of the $X(4260)$
 propagator.

Beside the $J/\psi\pi\pi$ channel indicated by the experiment, the
X(4260) may also decay into $h_c\pi\pi$,and there are other nearby
thresholds close to X(4260), such as $\chi_{c0}\,\omega$ (4197MeV),
$D_s^{*+}\bar D_s^{*-}$ (4224MeV), $D^-D_1^+(2420)$ (4291MeV),
$\chi_{c1}\,\omega$ (4293MeV),  etc.. It is possible that X(4260)
couples to all these channels. Therefore a careful way is to write
down the denominator of the $X(4260)$ propagator as:
  \bea\label{X4260Den}
  D_X(q^2)=M_X^2-q^2-i\sqrt{q^2}\Gamma(q^2),
  \eea
where $\Gamma_X(q^2)$ consists of all partial widths,
\begin{eqnarray}\label{gammax}
  \Gamma_{X}(q^2)=\Gamma_{J/\psi\pi\pi}(q^2)+\Gamma_{h_c\pi\pi}+g_{\omega\chi_{c0}}k_{\omega \chi_{c0} }+g_{D_s^*D_s^*}k_{D_s^*D_s^*}^3+g_{DD_1}k_{DD_1}+g_{\omega\chi_{c1}}k_{\omega\chi_{c1}}+\Gamma_0.\no\\
\end{eqnarray}
Here $\Gamma_{J/\psi\pi\pi}(q^2)$ is calculated from the above amplitude of $X(4260)\rightarrow J/\psi\pi\pi$,
 and $k_{\omega \chi_{c0}}$, $k_{D_s^*D_s^*}$, $k_{DD_1}$, $k_{\omega \chi_{c1} }$ are the 3-momentum
 of $\omega \chi_{c0},D_s^*D_s^*, DD_1,\omega \chi_{c1}$ in the $X(4260)$ rest frame, respectively.
 The $D_s^*D_s^*$ channel begins with a P-wave coupling therefore it depends on 3rd-order momentum,
 and the other three channels are of S-wave couplings and hence only depend on their 1st-order momentum.
 The possible rest partial widths are parameterized as a constant $\Gamma_0$.
For $h_c\pi\pi$, since the channel momentum $k_{h_c\pi\pi}$ behaves
like a constant near $q^2=M_X^2$, we parameterize
$\Gamma_{h_c\pi\pi}$ as a constant, too. Notice that
$\Gamma_{h_c\pi\pi}/\Gamma_{J/\psi\pi\pi}(M_X^2)$ is
constrained by experiment. We discuss this point in more detail
in the next section.


{{One may notice that $Z_c\pi$ is also a possible decay channel of
the X(4260) but it is not considered in this work, since the
contribution of $X(4260)\to Z_c\pi\to J/\psi\pi\pi$ can be absorbed
into the $XJ/\psi\pi\pi$ contact interaction in the Lagrangian
Eq.\,(\ref{XJPsipipi}). 
To confirm this viewpoint, we tested the
contribution of $Z_c\pi$ with Breit-Wigner parametrization of $Z_c$,
and found that there was not much difference in $J/\psi\pi\pi$ and
$\pi\pi$ spectrums from that without its contribution.  Therefore we
exclude $Z_c\pi$ contribution in this paper and leave it for future
work.}}


In the present work we fit the X(4260) line shape in the region of
4.15GeV$<\sqrt{q^2}<4.47$~GeV, where the data is from
Ref.~\cite{Belley4260} (16 data points) and Ref.~\cite{BaBar2012}
(16 data points), see in Figure~\ref{1234invariant mass}a. The total
cross section of $e^+e^-\rightarrow J/\psi\pi\pi$ is given by
\begin{eqnarray}\label{cs}
\sigma_{e^+e^-\rightarrow J/\psi\pi^+\pi^-}&=&\int_{s_-}^{s_+}\int_{t_-}^{t_+}\frac{\overline{|\mathcal{A}_1|}^2 d s d t}{(2\pi)^3 32 (q^2)^2}\,,
\end{eqnarray}
where $t=(q-q^-)^2$, $\mathcal{A}_1$ is defined in Eq. \ref{MP} and the lower and upper limits are given as
\begin{eqnarray}\label{intlimits}
s_- &=& 4 m_{\pi}^2 \, , \nonumber \\
s_+ &=& (\sqrt{q^2}-M_\Psi)^2 \, , \nonumber \\
t_{\pm} &=& \frac{1}{4\, s} \left\{ \left( q^2-M_\Psi^2 \right)^2 - \left[ \lambda^{1/2}(q^2,s,M_\Psi^2) \mp \lambda^{1/2}(s,m_{\pi}^2,m_{\pi}^2)\right]^2 \right\} \, ,\\
\lambda(&a,&b,c)=(a-b-c)^2-4bc\, . \no
\end{eqnarray}
With Eq.~(\ref{intlimits}), one finds out that for a larger $q^2$,
the upper limit of $s$ becomes too large to insure the validity of
the parametrization introduced in section 2.3\footnote{For
$\sqrt{q^2}=4.47$GeV, it corresponds to a value $\sqrt{s}\simeq
1.17$GeV which is within the range that $T$ can provide a reasonable
description.}. For $\pi\pi$ invariant mass spectrum we use the data
given in figure 4b of Ref.~\cite{Belley4260}, corresponding to
$\sqrt{q^2}\in [4.2, 4.4]$GeV (17 data points), the  data $\sqrt{q^2}\in [4.15,4.45]$GeV (41 data points) in
Ref.~\cite{BaBar2012}, and the  data
$\sqrt{q^2}$ fixed at 4.26GeV(44 points) from the recent
experiment~\cite{13035949}.
 For the first set of data, a MC study of
efficiency correction at $\sqrt{q^2}=4.26$GeV is given in Ref.~\cite{Belley4260}, and through a numerical
test we find that the efficiency curve
 is well reproduced by the $\pi\pi$ two body phase space up to a normalization constant,
 hence in our fit we simply use the two body phase space instead of
 the efficiency corrected one. We assume the other two sets of data maintain similar
 behavior. Summing up,  there are totally 145 data
points in $\pi^+\pi^-$ invariant mass spectrum to be used.

Recently the cross section of $\omega\chi_{c0}$ channel is measured
in $[4.21,4.45]$GeV~\cite{omegachic0scp,Ablikim:2014qwy}, where a rough structure of
the $X(4260)$ could be observed at 4.26GeV. If we assume the events
in Ref.~\cite{omegachic0scp,Ablikim:2014qwy} all come from $X(4260)$ decay,  then 9
data points can be used. Nevertheless it may be possible that these
events come from continuum rather than from $X(4260)$, hence in our
fit we carefully consider this possibility. That is, in the
following section~\ref{w/o}, the fit named Fit I, does not include
the $\omega\chi_{c0}$ cross section data, while in
section~\ref{with} the $\omega\chi_{c0}$ cross section data are
fitted, named Fit II. It is found that Fit I does not differ much
from the result of V1 (notice that here we include the $h_c\pi\pi$
channel), but Fit II, with the $\omega\chi_{c0}$ data being included,
violates the approximate `scaling law' found in V1 and lead to a
small value of $\Gamma_{e^+e^-}$.

Parameters needed in our fit are the following: Firstly,
Eq.~(\ref{eq1}) and (\ref{XJPsipipi}) describing the $\gamma$ -- $X$
transition and the tree level $X(4260)$ $J/\psi$ $\pi\pi (K\bar K)$
interactions provide 4 parameters. Secondly, it is found that taking
$\alpha_{1,2}(s)$ to be linear polynomials (except the Adler zero
term) is already good enough for data fitting, hence the two
$\alpha_i(s)$ (i=1,2 and each $\alpha_i(s)$ contains two parameters)
contribute another 4 parameters. Thirdly the  mass $M_X$ of the
$X(4260)$ in the propagator and the coupling constant
$g_{\omega\chi_{c0}}$ bring another two parameters. Finally, there
are three normalization factors $N_1$, $N_2$ and $N_3$ for the
$\pi^+\pi^-$ invariant mass spectrum of Belle, Babar and BESIII,
respectively. After summing up there are totally 13 parameters.

We also studied the parameters $g_{D_s^*D_s^*},g_{DD_1}$,
$g_{\omega\chi_{c1}}$ and $\Gamma_0$ through rather extensive
numerical tests in different environments, and we found that the
$g_{D_s^*D_s^*},g_{DD_1}$ and $g_{\omega\chi_{c1}}$ are always
vanishingly small which suggests that the coupling of X(4260) to
$D_s^*D_s^*$, $DD_1(2420)$ and $\omega\chi_{c1}$ are negligible,
compared to $J/\pi\pi$, $\omega\chi_{c0}$ and $h_c\pi\pi$. Moreover,
the parameter $\Gamma_0$ tends to vanish in all different fits and
hence it can be ignored too.\footnote{ It is exactly for this reason
we will band the $h_c\pi\pi$ width with the $J/\Psi\pi\pi$ width
with a ratio $R$ (see Eq.~(\ref{Gammahcpipi})),  otherwise the fit
program will tend to destroy it. } Therefore we will not include the
couplings to $D_s^*D_s^*$, $DD_1(2420)$, $\omega\chi_{c1}$ and
$\Gamma_0$ in our discussion from now on.


\subsection{Fit without $\omega\chi_{c0}$ Data}\label{w/o}

In this fit, named as Fit I, only the $J/\psi\pi\pi$ cross section
and $\pi\pi$ invariant mass spectrum are included, but not the
$\omega\chi_{c0}$ cross section data which will be analyzed in the
next section. 
About $h_c\pi\pi$ channel, however, on one hand,
lacking of precision the rough shape is not appropriate to fit \cite{Yuan:2013ffw}.
On the other hand, it is not clear yet whether they are from the
$X(4260)$ resonance or from the continuous background.  Therefore we
assume $h_c\pi\pi$ has an unknown width in the $X(4260)$ propagator,
and it is assumed to be a constant because its threshold is far away
from the $X(4260)$ resonance,\footnote{In fact, the
momentum-dependent width of $h_c\pi\pi$ channel with form factor is
also tested in the fitting as following, \bea\label{threebody}
\Gamma_{h_c\pi\pi}(q^2)\sim\int\frac{\sqrt{(q^2-(\sqrt{s}+m_{h_c})^2)(q^2-(\sqrt{s}-m_{h_c})^2)}\sqrt{s-4m_{\pi}^2}}{4qs}dsd\Omega,
\eea and \bea
\Gamma_{X}(q^2)=(\Gamma_{J/\psi\pi\pi}(q^2)+\Gamma_{h_c\pi\pi}(q^2))\exp(\frac{q^2-M^2_{X}}{\Lambda^2})+g_{\omega\chi_{c0}}k_{\omega \chi_{c0} }+\cdots.\no\\
 \eea
This fitting result is always similar with the constant width of $h_c\pi\pi$, therefore only the constant $\Gamma_{h_c\pi\pi}$ is shown in the text.} and it is constrained by the width of $\Gamma_{J/\psi\pi\pi}(q^2)$ as the following relation
\bea\label{Gammahcpipi}
 \Gamma_{h_c\pi\pi}=R \times\Gamma_{J/\psi\pi\pi}(q^2)|_{q=4.26\mbox{GeV}},
\eea where the coefficient $R=0.66$ is a rough estimation  from the
ratio of $\sigma_{h_c\pi\pi}/\sigma_{J/\psi\pi\pi}$ at
$q=4.26$GeV\cite{Ablikim:2013wzq}. Of course, we will also test the
fits with different $R$ value ranging from 0 to 0.66.

\begin{figure}[htp]
\centering
    \includegraphics[height=2.5in,width=3.2in]{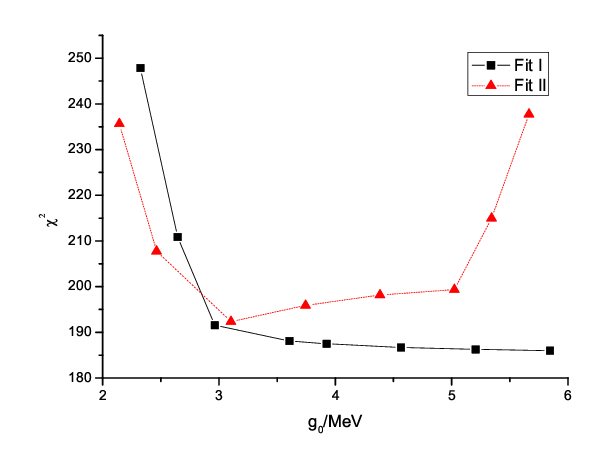}
 \caption{\label{chi2} $\chi^2$ dependence on $g_0$. Fit I: without the $\omega\chi_{c0}$ data, where  we see the scaling behaviour; Fit II: with the $\omega\chi_{c0}$ data where the scaling behaviour disappears.}
\end{figure}

\begin{figure}[htp]
\centering
\subfigure[]{
    \includegraphics[height=2.1in,width=2.8in]{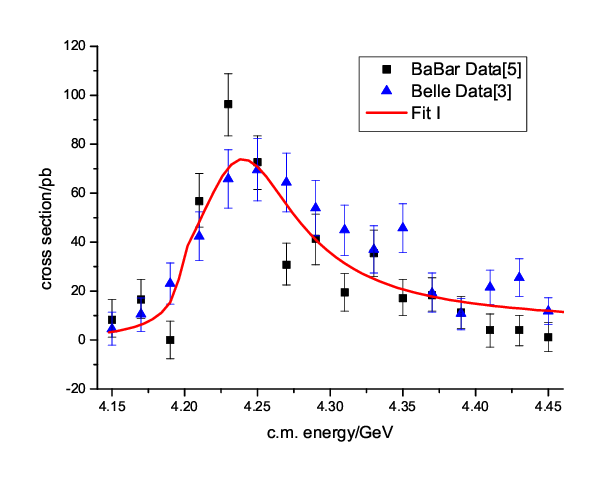}}
\subfigure[]{
    \includegraphics[height=2.1in,width=2.8in]{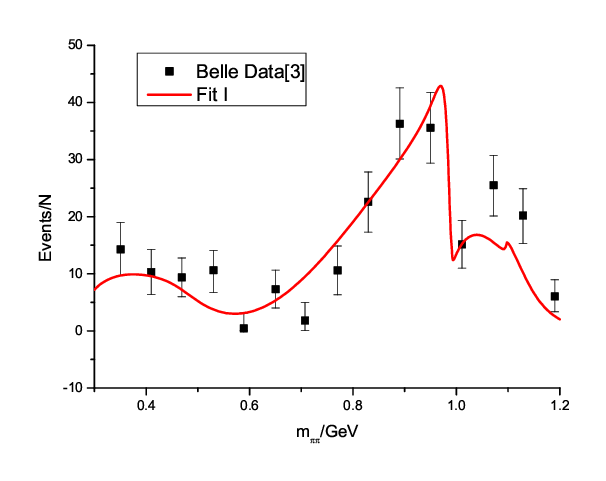}}
\subfigure[]{
    \includegraphics[height=2.1in,width=2.8in]{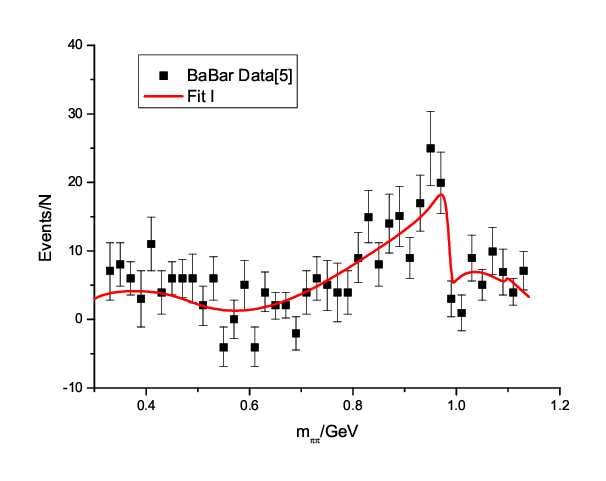}}
\subfigure[]{
    \includegraphics[height=2.1in,width=2.8in]{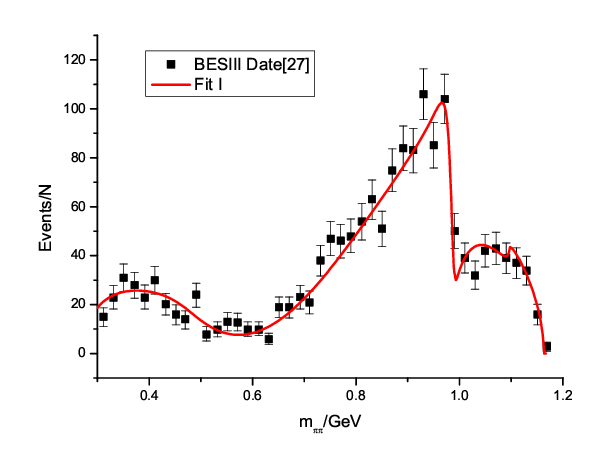}}
 \caption{\label{1234invariant mass} The fit to the cross section $e^+e^-\rightarrow J/\psi\pi\pi$ and $\pi\pi$
 invariant mass spectrum  without $\omega\chi_{c0}$ data.}
\end{figure}

\begin{table}[htp]
\begin{center}
\begin{tabular}{c|c|c}
\hline
                            &  Fit I                         &    Fit II                \tabularnewline
\hline
$\chi^{2}/d.o.f.$    & 187.1/(145-12)        & 193.5/(154-13)       \tabularnewline
\hline
$g_{0}(\mbox{MeV})$ &  $4.24$ (fixed)   &  $3.32\pm0.11$     \tabularnewline
$g_{Y\omega\chi_{c0}}$  & $0.17\pm0.01$   & $0.06\pm0.01$   \tabularnewline
$M_{X}(\mbox{GeV})$     &  $4.2504\pm0.0034$  &  $4.2432\pm0.0031$  \tabularnewline
\hline
\end{tabular}
 \caption{\label{fit1234} Parameters from Fit I and Fit II. Ratio R is chosen at $0.66$ for example.
 Since $g_0$ is fixed,  the total parameters are 12, and only important parameters are presented. }
\end{center}
\end{table}

When fit to the $J/\psi\pi\pi$ cross section and the $\pi\pi$
invariant mass, we found that the value of $g_0$ has large
uncertainty. Since it characterizes the transition between $X(4260)$
and the photon field, or $\Gamma_{X\to e^+e^-}$ according to
Eq.~(\ref{Gammaee}), it should not be too large to avoid the upper
bound established  by an analysis on the BES experiment:
Ref.~\cite{moxh} gives $\Gamma_{X\to e^+e^-}<420$~eV, or most
conservatively $<580$~eV, at $90\%$ confidence level. The dependence
 of $\chi^2$ on $g_0$ is clearly depicted in
Fig.~\ref{chi2} (black square) which exhibits an approximate scaling
law on parameter $g_0$. When $g_0$ increases, parameters $h_i$ have
to become small to compensate for the experimental value of
$\Gamma_{X\to J/\Psi\pi\pi}$ as given in Eq.~(\ref{XJPsipipi}). See
Eqs.~(\ref{eq1}) and (\ref{A}), it is easy to find out that
$\Gamma_{X\to J/\Psi\pi\pi}$ is proportional to the product of $g_0$
and $h_i$.
This mechanism keeps $\Gamma(q^2)$ of the denominator of $X(4260)$
propagator (see Eq.~(\ref{X4260Den})) almost unchanged when $g_0$
varies. As a consequence, $\chi^2$ becomes almost inert with respect
to the variation of $g_0$ when it is large enough since the effect
can be counterbalanced by a variation of $h_i$. This observation is
already made in V1, here we reconfirm the `scaling law' even when
$\Gamma_{h_c\pi\pi}$ is included.

The `scaling law' means that we can not reliably determine parameter
$g_0$ at all when $g_0$ is large enough. It is important to notice
that in Fig.~\ref{chi2} there exists a large enough space for $g_0$
to maintain a (almost) minimal $\chi^2$, and at lower value it is
below  the BES $\Gamma_{X\to e^+e^-}$ bound given
in~Ref.~\cite{moxh}. As an example, we list the lowest value of
$g_0$ at 4.24MeV which corresponds to $\Gamma_{X\to e^+e^-}\simeq
41.1$~eV, and its fitting result is shown in
Fig.\,\ref{1234invariant mass} and Table\,\ref{fit1234}. For this
chosen value, Table.\,\ref{fit1234} indicates the widths of
$J/\psi\pi\pi$, $h_c\pi\pi$ and $\omega\chi_{c0}$ at
$\sqrt{q^2}=4.26$~GeV are 32.4~MeV, 21.3~MeV and 49.8~MeV,
respectively.

 We also search for poles of the $X(4260)$ propagator, and the complex plane is divided into four Riemann sheets
 by $\Gamma_{J/\psi\pi\pi}$ ($\Gamma_{h_c\pi\pi}$) and $\Gamma_{\omega\chi_{c0}}$ defined in Table\,\ref{Riemannsheet},
 and the pole positions are presented in Table\,\ref{fitIpole}. It is noticeable that there are two pair of poles
 locating above the $\omega\chi_{c0}$ threshold on the third and fourth sheet, and both of them show around 85 MeV
 pole width for the X(4260) resonance. According to the pole counting rule in Ref.\,\cite{polecounting,zhangoux3872},
 the two pair of near threshold poles indicate that $X(4260)$ is not like a molecule resonance but more like
 an
`elementary' particle or, in other words, a confining state.

  \begin{table}[t!]
\begin{center}
 \begin{tabular}  {c| c c c c}
 \hline\hline
           &   sheet I    &   sheet II     &   sheet III    & sheet IV  \\ \hline
$\Gamma_{J/\psi\pi\pi}+\Gamma_{h_c\pi\pi} $   &  +           &  -             &   -            &  +    \\
$\Gamma_{\omega\chi_{c0}}$    &  +           &  +             &   -            & -             \\
\hline\hline
 \end{tabular}
 \caption{\label{Riemannsheet} {\small
  Definition of the four Riemann sheets with
 the $J/\psi\pi\pi$ ($h_c\pi\pi$) channel and $\omega\chi_{c0}$ channel. }}
   \end{center}
\end{table}

   \begin{table}[t!]
\begin{center}
 \begin{tabular}  {c|clclclc}
 \hline\hline
            &  sheet I    &   sheet II     &   sheet III    & sheet IV  \\ \hline
   Fit I &           --           &  --             &   4231.9-44.2i            &  4233.2-42.5i        \\
\hline
  Fit II &            --           &  4241.5-24.4i             &   4232.8-36.3i            & -- \\
\hline
 \end{tabular}
 \caption{\label{fitIpole} {\small
Pole position of Fit I and II. The value of ${  \sqrt{s_{\rm pole}}=M_{\rm pole} - i\Gamma_{\rm pole}/2 }$
 is given in MeV.}}
   \end{center}
\end{table}

For the purpose in extracting solid physical conclusions, we also
varies the coefficient R as $0.56,\,0.46,\,0.36,\,0.26,\,0.16,\,0$
 in the fit,  and when $R=0$, the fit procedure is
very similar with V1\cite{Dai:2012pb}, where we found a large
coupling constant $g_{\omega\chi_{c0}}$ and also the $\chi^2$
scaling law on $g_0$. It is found that qualitative physical results are not
sensitive to the $R$ value, and all the fits with different $R$ maintain the same scaling law
on the coupling constant $g_0$. In all cases the large
coupling of the $\omega\chi_{c0}$ to the X(4260) always exists which
coincides with the  prediction in V1.

 Before going to the next section, a brief conclusion is in order: the fit result shows that the
  $\chi^2$ has an approximate scaling law on $g_0$, which could not be determined very well.
  Two nearby poles are found on the
  third and fourth Riemann sheet. The fit results are not sensitive to the R value,
  {which indicates that $h_c\pi\pi$ only contributes to the X(4260) decay as a smooth
  background.}
  We predict that the $\omega\chi_{c0}$ channel has a large branching ratio in the X(4260) decay in all cases.

 \subsection{Fit Including $\omega\chi_{c0}$ Cross Section Data}
\label{with}
 In the above subsection, we confirmed that there should be
a sizable contribution from $\omega\chi_{c0}$ channel even when the
recent $\omega\chi_{c0}$ cross section data from BESIII
collaboration\cite{omegachic0scp,Ablikim:2014qwy} are not  taken
into consideration. In this subsection, we further include the
$\omega\chi_{c0}$ cross section data in our analysis to get a more
precise result of $g_{\omega\chi_{c0}}$. This will certainly benefit
our understanding on $X(4260)$, provided that those
$\omega\chi_{c0}$ data indeed come from $X(4260)$ decay.  The cross
section of $e^+e^-\to X(4260)\to\omega\chi_{c0}$ is parameterized as
Ref.~\cite{PDG}, \bea\label{crosssection1}
  \sigma_{e^+e^-\rightarrow X(4260)\rightarrow \omega\chi_{c0} }(q^2) =
   \frac{3\pi}{4q^2}\frac{\Gamma_{ee}\Gamma_{\omega\chi_{c0}}}{|D_X(q^2)|^2},
\eea where $\Gamma_{ee}$ is given in Eq.\,(\ref{Gammaee}),
$D_X(q^2)$ is the denominator of the $X(4260)$ propagator shown in
Eq.\,(\ref{X4260Den}) and (\ref{gammax}) and
$\Gamma_{\omega\chi_{c0}}=g_{\omega\chi_{c0}}k_{\omega \chi_{c0}}$.

The major difference  after taking the $\omega\chi_{c0}$ data into
Fit\,II, is that the $\chi^2$ scaling law on $g_0$ disappears. As
shown in Fig\,\ref{chi2} (green triangle), the $\chi^2$ has the
minimum at $g_0\simeq3.32$~MeV. It suggests that this fit is
more stable than Fit\,I without the $\omega\chi_{c0}$ cross section.

\begin{figure}[htpd]
\centering
\subfigure[]{
    \includegraphics[height=2.1in,width=2.8in]{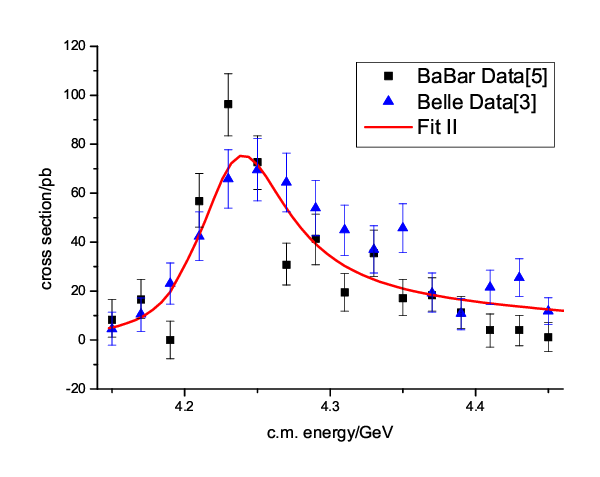}}
\subfigure[]{
    \includegraphics[height=2.1in,width=2.8in]{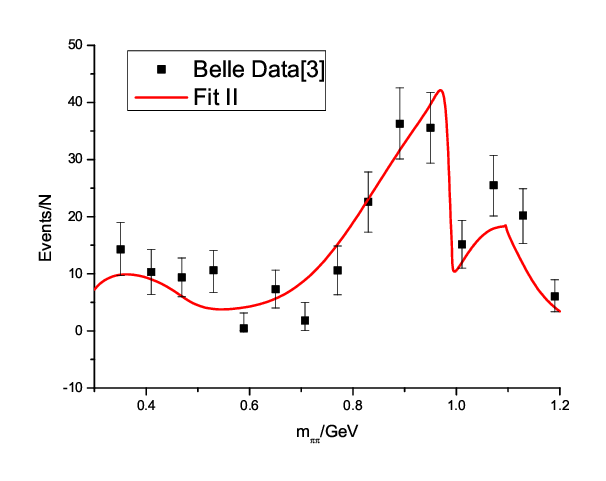}}
\subfigure[]{
    \includegraphics[height=2.1in,width=2.8in]{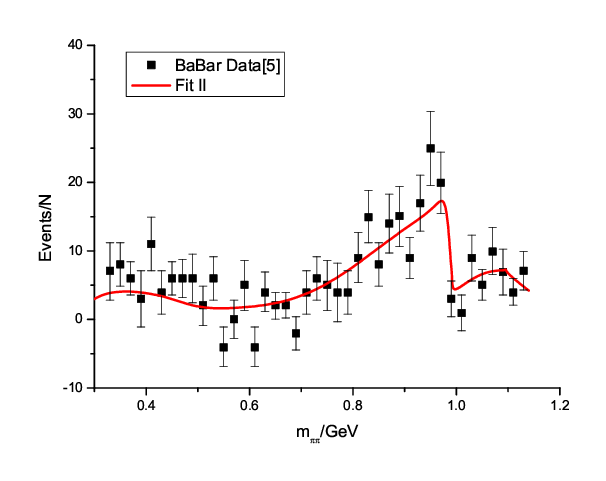}}
\subfigure[]{
    \includegraphics[height=2.1in,width=2.8in]{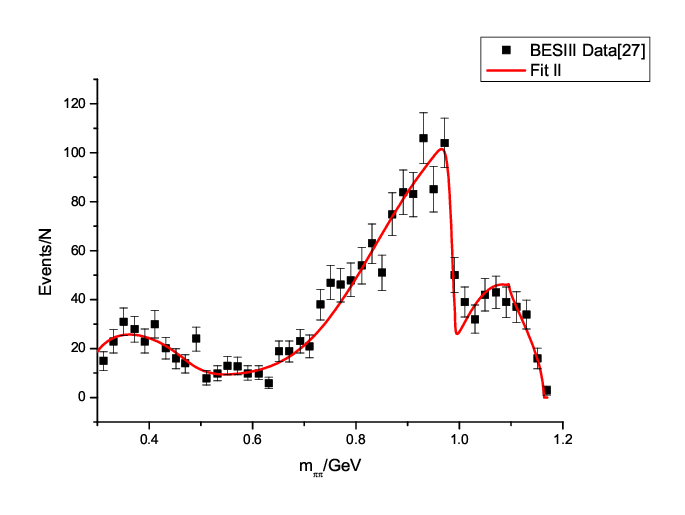}}
\subfigure[]{
    \includegraphics[height=2.1in,width=2.8in]{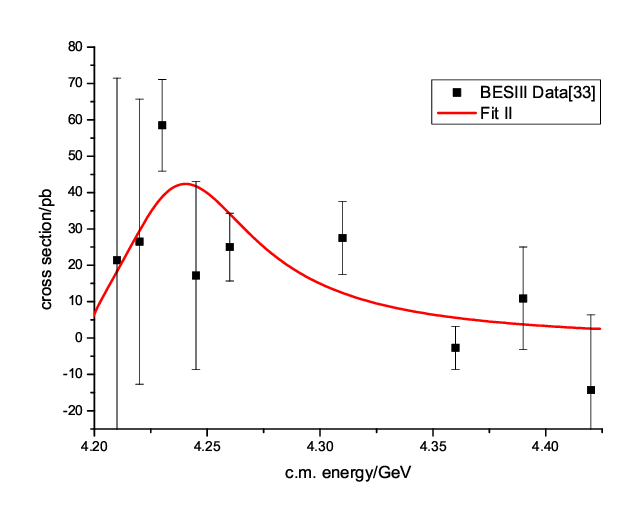}}
 \caption{\label{56invariant mass} The cross section of $e^+e^-\rightarrow J/\psi\pi\pi$, invariant mass of $\pi\pi$, and the cross section of $e^+e^-\rightarrow \omega\chi_{c0}$} 
\end{figure}

We again take  $R=0.66$ as the example,  the fit result is shown
in Fig.\,\ref{56invariant mass} and Table\,\ref{fit1234}. From
Table.~\ref{fit1234}, one notices that the coupling of
$\omega\chi_{c0}$ approaches a smaller value (but still much larger
than  other open-charm channels), which is well constrained by the
new data of X(4260) decaying into $\omega\chi_{c0}$.
The pole positions are also searched for and are shown in
Table\,\ref{fitIpole}. It should be noticed that there are two
pairs of poles on sheet II and III, which has smaller pole width
(around 60 MeV) comparing with Fit I.

Comparing  the $\chi^2/dof$, Fit\,II (193.5/141) is similar in quality
 with that of Fit\,I (187.1/133). The major different physical
impact once the $\omega\chi_{c0}$ data are included is that it
roughly reduces $g_{\omega\chi_{c0}}$ by a factor 3. The three
partial widths $\Gamma_{J/\psi\pi\pi}$, $\Gamma_{h_c\pi\pi}$ and
$\Gamma_{\omega\chi_{c0}}$ are now $45.1$MeV, $29.9$MeV, $16.9$MeV
at $\sqrt{q^2}=4.26$GeV, respectively. The branching ratio of
$\omega\chi_{c0}$ in this fit is $18.5\%$, which is still sizable,
and is still much larger comparing with  open-charm channels, such
as $D\bar D$, $D^*\bar D^*$ and $D_s^*\bar D_s^*$. Furthermore, we
{verified}  that the qualitative result is  not sensitive to
the ratio $R$ ranging in [0,\,0.66]. See Table~\ref{Rbehaviour} for illustration.
   \begin{table}[htpd]\label{Rbehaviour}
\begin{center}
\begin{tabular}{c|c|c|c|c|c|c|c}
\hline
 & \multicolumn{7}{c}{Fit II}\tabularnewline
\hline
R& 0.66 & 0.56 & 0.46 & 0.36 & 0.26 & 0.16 & 0.00 \tabularnewline
\hline
$\Gamma_{ee}\quad\ \ $ & 25.23 & 24.36 & 23.55 & 22.82 & 22.19 & 22.03 & 22.02  \tabularnewline
\hline
$\Gamma_{\omega\chi_{c0}}\ \ $ & 17.49 & 19.04 & 20.89  & 23.11  & 25.84  & 26.75 & 26.77  \tabularnewline
\hline
$\Gamma_{J/\psi\pi\pi}$& 54.52 & 57.02 & 59.67 & 62.42 & 65.21 & 65.82 & 65.02 \tabularnewline
\hline
$\Gamma_{h_{c}\pi\pi}$\ \ \ & 35.99 & 31.93 & 27.45 & 22.47 & 16.95 & 10.53 & 0   \tabularnewline

\hline
\end{tabular}
 \caption{ Partial widths obtained when R changes.
 The unit of $\Gamma_{ee}$ is in eV and the unit of the others is in MeV.}
   \end{center}
\end{table}

For a summary, after taking the $\omega\chi_{c0}$ cross section data
into account, the coupling of the X(4260) decay into
$\omega\chi_{c0}$ becomes smaller but fixable, and even though it is
no longer dominant but still plays an important role in the
$X(4260)$ decay, which supports our conclusion before. Certainly,
because of the large error bar near 4.26GeV from the
$\omega\chi_{c0}$ cross section data (in Fig.~\ref{56invariant
mass}(e)) a more qualitative  conclusion still needs more statistics
from the experimental data.


\section{Conclusions and Discussions}\label{conclusion}
The property of $X(4260)$ remains mysterious after being
discovered for almost ten years.  The nature of this particle is
still an controversial issue. In this work, we investigate this
particle based on all experimental data available and very modest
theoretical assumptions. Hence the conclusion we reached should be
 robust. Comparing with V1~\cite{Dai:2012pb}, two more
channels $h_c\pi\pi$ and $\omega\chi_{c0}$ are now considered.

  We have performed two fits: Fit\,I without and Fit\,II with the $\omega\chi_{c0}$ cross section data.
  In the two scenarios, they have similar $\chi^2/dof$ but different behavior on the coupling constant $g_{0}$.
In the former one a $\chi^2$ scaling law on $g_0$ is observed which could not be determined from the fit,
  while in the latter case the $\chi^2$ scaling law on $g_0$ disappears which has a minimum value when  $g_0=3.32$MeV.  The value of $g_0$ corresponds to $\Gamma_{e^+e^-}\simeq25$eV, which is certainly
  below
  the $\Gamma_{e^+e^-}$ bound in BES experiment~\cite{moxh}. Considering the variation of R, we also give the following estimate on \bea\Gamma_{ee}=23.30\pm3.55 \text{eV}\ .\eea Compared with the experimental observation of $\Gamma_{ee}\times\mathrm{Br}(J/\psi\pi\pi)=9.7\pm1.1$eV\cite{Belley4260} or $9.2\pm1.5$eV\cite{BaBar2012}, we conclude that the roughly half of $X(4260)$ decay into $J/\psi\,\pi\pi$ final state.
  {Finally, our
analysis points out a sizable coupling between $X(4260)$ and
$\omega\chi_{c0}$ which awaits a theoretical explanation.}

  In both fits there
are two nearby poles found in X(4260) propagator indicating
that the $X(4260)$ is most likely a confining state~\cite{polecounting, zhangoux3872}.
   The small value of $\Gamma_{e^+e^-}$ is consistent with the hybrid
   scenario which indicates
 $
 5.5\pm1.3\mbox{eV}\le \Gamma_{e^+e^-} \le 62 \pm 15\mbox{eV}$~\cite{close05} or $23\pm20$eV~\cite{chengyin2014}.
Also the hybrid state is suppressed to decay into
$D\bar D$, $D_s\bar D_s$, $D^*\bar D^{*}$ and $D_s^*\bar
D_s^*$\cite{hybrid decay}, which coincides with the
experimental data for the $X(4260)$. Nevertheless, since $D_1$ is in $P$ wave, a hybrid state is likely to have a large coupling to $DD_1$ channel, this is not supported by experiment \cite{Ablikim:2013xfr} and our analysis.
It should be  pointed out that
 a small value of $\Gamma_{e^+e^-}$ is also consistent with the explanation that $X(4260)$ is the 3D charmonium state.
  The difficulty of this possible explanation comes from the role of $X(4160)$, which is considered as
   candidate of the 3D charmonium state in the literature, though it has a rather large $\Gamma_{e^+e^-}$ width.
 We hope our effort made in this paper will be helpful for future investigations in clarifying the issue of  $X(4260)$.

\section{Acknowledgement}
We are grateful to illuminating discussions with Chang-Zheng Yuan
and Kuang-Ta Chao, and would also like to thank Gui-Jun Ding, Ce
Meng, Qiang Zhao and Bing-Song Zou for helpful discussions.
 This work is supported in
part by National Nature Science
Foundations of China under contract number  10925522 
 and
11021092.

\renewcommand\refname{Reference}
\bibliographystyle{h-physrev}
\bibliography{4260version}

\end{document}